\documentclass[iop]{emulateapj}
\usepackage{amsmath, amssymb}
\usepackage{natbib}
\usepackage{graphicx}

\shorttitle{Magnificent Magnification}
\shortauthors{Huff \& Graves}

\begin{document}

\title{Magnificent Magnification: Exploiting the Other Half of the
  Lensing Signal}

\author{Eric M. Huff\altaffilmark{1,2} \& 
  Genevieve J. Graves\altaffilmark{1,2,3}}
\altaffiltext{1}{Department of Astronomy, University of California,
  Berkeley, CA 94720, USA}
\altaffiltext{2}{Lawrence Berkeley National Lab, 1 Cyclotron Road
  MS50R, Berkeley, CA 94720}
\altaffiltext{3}{Miller Fellow}


\begin{abstract}
  We describe a new method for measuring galaxy magnification due to
  weak gravitational lensing. Our method makes use of a tight scaling
  relation between galaxy properties that are modified by
  gravitational lensing, such as apparent size, and other properties
  that are not, such as surface brightness.  In particular, we use a
  version of the well-known fundamental plane relation for early type
  galaxies.  This modified ``photometric fundamental plane'' replaces
  velocity dispersions with photometric galaxy properties, thus
  obviating the need for spectroscopic data.  We present the first
  detection of magnification using this method by applying it to
  photometric catalogs from the Sloan Digital Sky Survey.  This
  analysis shows that the derived magnification signal is comparable
  to that available from conventional methods using gravitational
  shear.  We suppress the dominant sources of systematic error and
  discuss modest improvements that may allow this method to equal or
  even surpass the signal-to-noise achievable with shear.  Moreover,
  some of the dominant sources of systematic error are substantially
  different from those of shear-based techniques.  Thus, combining the
  two techniques addresses the major weaknesses of each and provides a
  substantial improvement over either method used in isolation.  With
  this new technique, magnification becomes a necessary measurement
  tool for the coming era of large ground-based surveys intending to
  measure gravitational lensing.
\end{abstract}

\section{Introduction}\label{sec:intro}
The modern cosmological concordance model has been spectacularly
successful.  This success has come at a price, however: cosmologists
must postulate a universe in which the vast majority of the content,
in the form of dark matter and dark energy, is inaccessible to direct
observation.

Gravitational lensing, and weak lensing in particular, provide our
best window onto the dark universe.  Because of this, the astronomical
community is investing heavily in current and future imaging surveys,
both ground- and space-based, designed at least in part around weak
lensing science, e.g., the Dark Energy Survey (DES), the Panoramic
Survey Telescope and Rapid Response System (Pan-STARRS), the Hyper
Suprime Cam for the {\it Subaru} telescope, the Large Synoptic Survey
Telescope (LSST), and Euclid.

Lensing measurements have already played a significant role in
astrophysics in the last two decades over a range of scales and
physical regimes.  Weak lensing measurements have characterized the
aggregate properties of galaxies' dark matter haloes, (e.g.,
\citealt{fischer00, sheldon04, hoekstra04, seljak05,
  mandelbaum06_halomass, parker07}), the dark matter profiles of large
galaxy clusters, both on a cluster-by-cluster basis (e.g.,
\citealt{kneib03, broadhurst05, clowe06_bullet, hoekstra07, jee07,
  mahdavi07_a520, mahdavi07_a478, berge08, bradac08, okabe08, kubo09})
and for stacked galaxy groups and clusters (e.g.,
\citealt{mandelbaum06_massprofiles};
\citealt{sheldon09, leauthaud10}) and, with recent cosmic shear
detections \citep{fu08,2007ApJS..172..239M}, directly measured the
clustering of matter on cosmological scales.

These measurements are currently made almost exclusively by studying
spatially-correlated distortions in the ellipticities of background
galaxies due to the shear component of the gravitational lensing
distortion.  In the weak-lensing regime, the ellipticities induced by
lensing are small ($\sim 1$\%) compared to the range of intrinsic
galaxy ellipticities ($\sim 30$\%) \citep{2005MNRAS.361.1287M}.

The lensing distortion also magnifies background sources, but the
intrinsic variance in the distribution of galaxy sizes and
luminosities---those properties perturbed by magnification---is much
larger than that of galaxy shapes. Magnification measurements to date
have necessarily had to average over much larger galaxy samples to
obtain signal-to-noise (S/N) equivalent to shear measurements
\citep{hildebrandt11,
  bauer11,2010MNRAS.405.1025M,2005ApJ...633..589S}.

In this Letter, we make use of a tight galaxy scaling relation---a
photometry-only version of the well-known fundamental plane for early
type galaxies \citep{djorgovski87, dressler87}---to substantially
narrow the intrinsic distribution of sizes for a set of background
source galaxies.  This makes it possible to measure the weak lensing
magnification signal around a set of foreground galaxy lenses with
higher S/N than was previously thought possible
\citep{2010arXiv1009.5735R}.  A thorough description of the
methodology and implementation of this analysis will be presented in a
forthcoming paper (hereafter Paper II). Here, we describe the core
concepts, present a detailed outline of the method, and demonstrate
its effectiveness.

In section \ref{sec:photoFP}, we describe the scaling relation used in
this analysis.  In section \ref{sec:mag_measurement}, we present a
proof-of-concept measurement of weak lensing magnification using this
technique and control for the most important systematic biases.  We
conclude in section \ref{sec:discussion} with a brief discussion of
the ways this method might be improved upon, with an eye toward
extracting a lensing signal from magnification that equals or even
exceeds the S/N obtainable from shear-based techniques.  Throughout,
we assume a $\Lambda$CDM cosmology with $\Omega_M = 0.274$,
$\Omega_{\rm{\Lambda}} = 0.726$, and $H_0 = 100$ km s$^{-1}$
Mpc$^{-1}$.

\section{The Photometric Fundamental Plane}\label{sec:photoFP}
The fundamental plane (FP) is in many ways an ideal tool for measuring
magnification.  It is an observed correlation between galaxy effective
radius ($R_e$), which is magnified by gravitational lensing, and two
galaxy properties which are unaltered by lensing: galaxy surface
brightness ($\mu$) and the stellar velocity dispersion ($\sigma$).
The intrinsic scatter in the FP is $\sim 0.08$ dex \citep{jorgensen96,
  bernardi03c}, or 20\%.  Thus the FP makes it possible to
predict the intrinsic value of $R_e$ from observations of $\mu$ and
$\sigma$, which can then be compared with the observed values of $R_e$
to measure magnification.

The FP was in fact proposed as a tool for this purpose by
\citet{bertin06}, but to our knowledge has never been used as such due
to a critical flaw.  Placing galaxies on the FP requires $\sigma$
measurements. Even with the tight scatter in the FP, a statistically
viable measurement would require high-resolution spectroscopic
measurements for millions of galaxies.

Identifying a purely photometric analog to the FP with comparable
scatter would solve this problem. Such a relation has already been
identified by \citet{graham02}, where the concentration of the galaxy
light profile fills the role normally served by $\sigma$. This works
in part because concentration and velocity dispersion are both
strongly correlated with galaxy mass, and in part because at fixed
mass, galaxies with more concentrated mass profiles have higher
velocity dispersions. We will explore the relation between the
spectroscopic fundamental plane and the photometric relation deployed
here more fully in Paper II.

\subsection{Background Sources}\label{sec:sources}
To define a photoFP for this work, we use a sample of galaxies drawn
from the Sloan Digital Sky Survey III (SDSS-III) Eighth Data Release
(DR8, \citealt{aihara11}).  We limit the sample to resolved sources
that meet basic quality cuts (e.g., are not saturated).  For these, we
estimate photometric redshifts (photo-$z$'s) based on the SDSS {\it
  ugriz} photometry using the public code ZEBRA \citep{feldmann06} run
with the default templates, allowing interpolation between the
standard templates without template optimization.  To select a sample
of early type background sources that should lie on the photoFP, we
exclude the $\sim 2/3$ of the galaxies with best-fitting templates
inconsistent with that of a passive stellar population. The sample
selection for background sources will be described in greater detail
in Paper II.

\begin{figure}[t]
\includegraphics[width=1.0\linewidth]{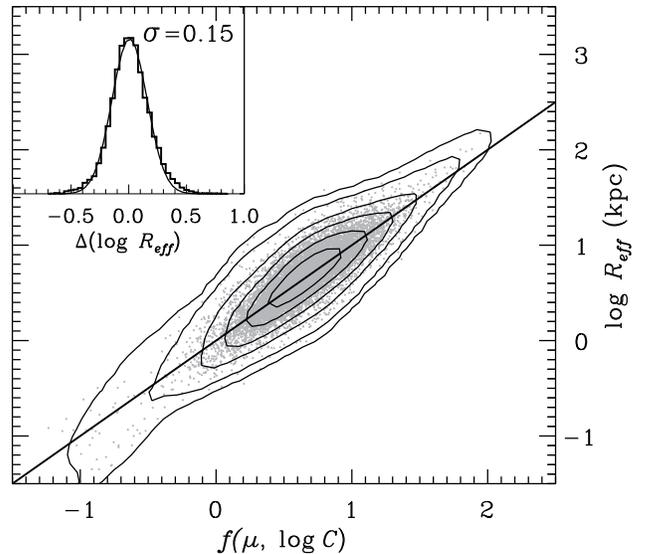}
\caption{The photometric fundamental plane for our source sample of
  8.4 million galaxies, shown edge-on.  $\log R_e$ is fit as a
  function of effective surface brightness ($\mu$) and concentration
  ($\log C$) separately in redshift bins of width $\Delta z = 0.01$.
  Gray points show a random subset of 100,000 galaxies from the source
  catalog, while the solid line shows the one-to-one relation.
  Contours enclose the 0.5$\sigma$, 1$\sigma$, 1.5$\sigma$, 2$\sigma$,
  2.5$\sigma$, and 3$\sigma$ boundaries of the 2D distribution for the
  full source catalog.  The inset shows the distribution of residuals
  in $\log R_e$ from the photoFP fits, which has width $\sigma =
  0.153$ dex.  }
\label{fig:fp_edgeon}
\end{figure}

The SDSS photometric pipeline does not measure S{\'e}rsic index.
Here, we substitute for $n$ the SDSS petrosian concentration $C =
R\,\mbox{90}/R\,\mbox{50}$, defined as the ratio of the radii containing
90\% and 50\% of the Petrosian flux (e.g., \citealt{shimasaku01}).
All reported quantities are measured in the {\it r} band.

We fit a photoFP of the form
\begin{equation}\label{eq:photoFP}
\log R_e = \alpha \mu + \beta \log C + \gamma, 
\end{equation}
where $R_e$ is the half-light radius of the best-fit de Vaucouleurs
light profile converted into physical units using the ZEBRA photo-$z$,
$\mu$ is the mean de Vaucouleurs surface brightness within $R_e$, and
$\alpha$, $\beta$, and $\gamma$ are free parameters. To avoid errors
resulting from a redshift-dependent selection function, evolution in
the photoFP, and $K$-corrections to the radii due to the fact that
the morphological measurements are all made in the observed-frame {\it
  r} band, we divide our galaxy sample into redshift bins with width
$\Delta z = 0.01$ and fit the photoFP separately in each bin.  The
best-fit coefficients are chosen to minimize the dispersion in
effective radius at fixed $\mu$ and $\log C$, taking into account only
the errors in $R_e$.

Figure \ref{fig:fp_edgeon} shows an edge-on view of the photoFP for our
source sample.  The dispersion around the photoFP in the direction of
effective radius is 0.15 dex, or 35\%.  

\begin{figure}
\includegraphics[width=1.0\linewidth]{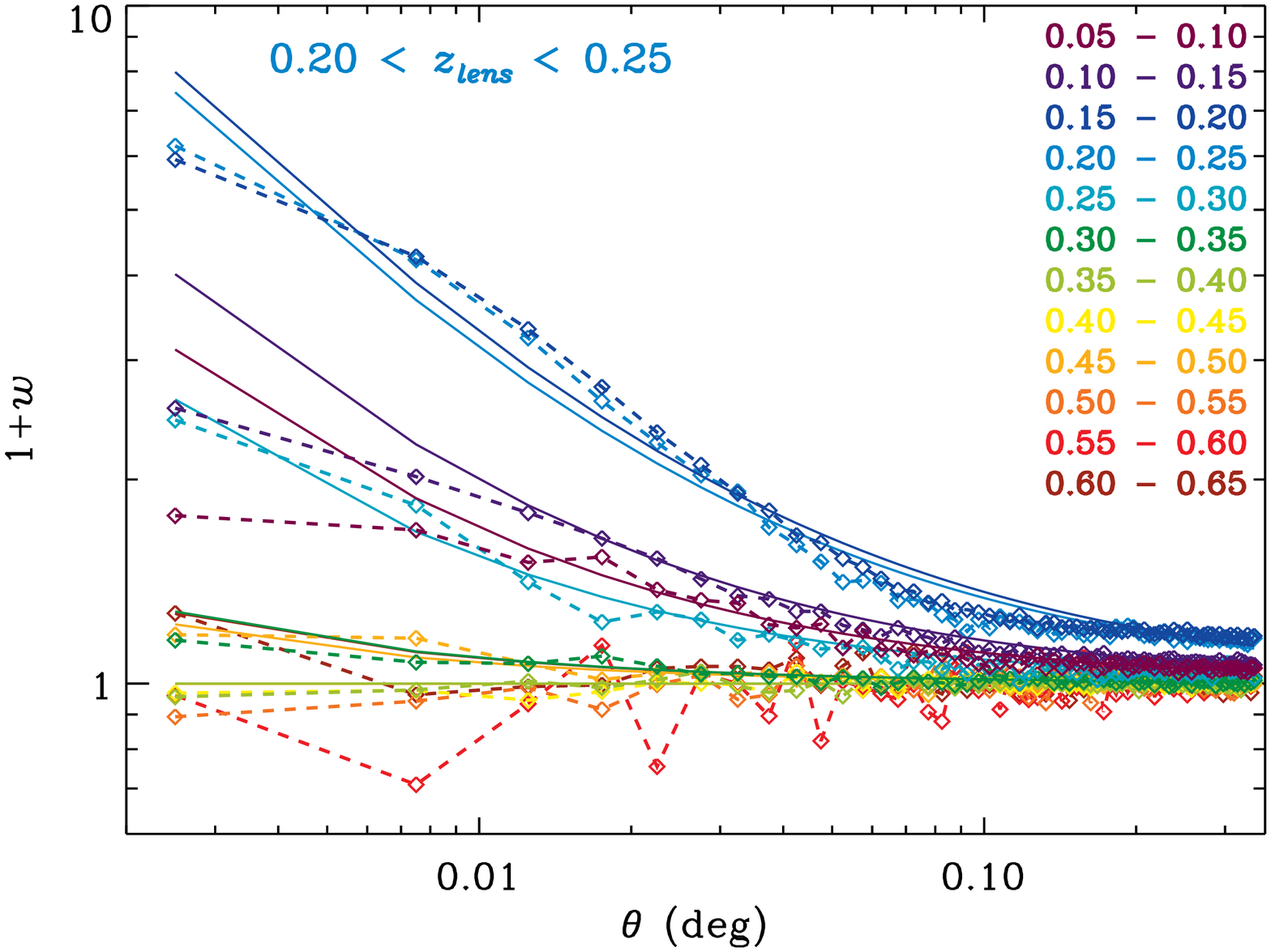}
\caption{The projected correlation function for sources
  around lenses with $0.20 < z_l < 0.25$.  Colors indicate different
  bins in $z_s$.  Solid lines show fits to the data.  At small
  separations, a large fraction of the galaxies in nearby $z$ bins are
  likely scattered in from $z_l$ through photo-$z$ errors.
  See text for details. }\label{fig:delta_resid}
\end{figure}

\subsection{Magnification using the photoFP}\label{sec:mag_photofp}

A line-of-sight matter overdensity at lens redshift $z_l$ will produce
an image convergence $\kappa$ of amplitude:
\begin{equation}\label{eq:kappa}
\kappa = \frac{\Sigma(d_l \vec{\theta})}{\Sigma_{\rm{crit}}},
\end{equation}
where $\Sigma$ is the projected surface density on the sky at $z_l$
and $\Sigma_{\rm{crit}}$ is the characteristic surface density of
matter required for lensing.  $\Sigma_{\rm{crit}}$ is defined by the
lensing geometry, such that
\begin{equation}\label{eq:sigma}
\Sigma_{\rm{crit}} = \frac{c^2}{4 \pi G} \frac{d_s}{d_l d_{ls}\left(1+z_l\right)^2},
\end{equation}
where $d_l$, $d_s$, and $d_{ls}$ are the angular diameter distances
from the observer to the lens, from the observer to the source, and
from the lens to the source, respectively. The factor of
$\left(1+z_l\right)^2$ arises from our use of comoving coordinates.

The lensing convergence re-scales the light profile, in the limit of
very weak lensing, by a factor of $\left(1+\kappa\right)$. The radius
and luminosity increase, but as the light profile is simply rescaled,
the concentration is left unchanged. In the presence of the scaling
relation described above, this implies an estimator ${\hat \kappa}$
of:

\begin{equation}\label{eq:kappa_est}
  \begin{split}
    \log \left(1+\hat{\kappa}\right) & = \Delta \log R_e \\
    & \equiv \log R_e - 
    (\alpha \mu + \beta \log C + \gamma).
  \end{split}
\end{equation}

If the errors in the observables are uncorrelated, the variance in our
estimator $ \hat{\kappa}$ is just the variance in the photoFP in the
direction of $R_e$. We extract a galaxy-galaxy lensing signal by
cross-correlating this estimator with a population of foreground
lenses.

\section{A Magnification Measurement}\label{sec:mag_measurement}

\subsection{Lens Sample}\label{sec:lenses}

The lens sample is selected from the NYU Value-Added Catalog
\citep{blanton05_vagc} version of the SDSS Data Release 7 (DR7)
spectroscopic survey \citep{abazajian09_dr7}, using only Luminous Red
Galaxy Sample targets (LRGs, \citealt{eisenstein03}).  In order to
compare with the results of \citet{mandelbaum08}, we limit the sample
to massive galaxies with absolute $r$-band magnitudes $-21.5 >
M_{^{0.0}r} > -22.6$ and redshifts $0.15 < z < 0.35$.  The magnitudes
are $k$-corrected and evolution corrected to $z=0.0$ as in
\citet[hereafter M+06]{mandelbaum06_massprofiles}. Finally, to exclude
satellite galaxies that are not at the centers of their dark matter
haloes, we remove galaxies with brighter nearby LRGs, again following
M+06 .  This gives a sample of $\sim 55,000$ lenses that have
comparable properties to the combined LRG sample of M+06.

\subsection{Correcting Biases due to Photometric Redshift
  Errors}\label{sec:pz_corr} 

In the presence of photo-z errors, the overdensity of sources
clustered near a lens will produce an excess of galaxies with
incorrect photo-z ($z_p$) along the line of sight to the lens. As a result, when
we average $\Delta \log R_e$ over the foreground or background source
galaxies, we systematically mis-estimate the residuals from the plane
associated with a lens due to the `shadow' cast by photo-z errors.

To deal with this bias, we will calculate the magnitude of this
spurious signal directly from the data, and subtract it from our
measured signal.  We must first estimate the error in $\Delta \log
R_e$ induced by a galaxy being assigned the wrong $z_p$ ($\Delta \log
R_e^{\rm{err}}$), then calculate what fraction $f_l$ of the galaxies
at each $z_p$ have been scattered in from $z_l$.  In these terms, the
observed mean photoFP residual is:
\begin{equation}\label{eq:kappa_pzerr}
  \Delta \log R_e^{\rm{obs}}= 
  \left(1-f_l\right)\log \left(1+ \kappa \right) + f_l \Delta \log R_e^{\rm{err}},  
\end{equation}
where $\kappa$ is the true convergence.

$\Delta \log R_e^{\rm{err}}$ can be estimated by assuming that the
galaxy lies on the photoFP at $z_l$ but is incorrectly assigned to
$z_p$.  The inferred effective radius of a galaxy with true redshift
$z_l$ that is mistakenly assigned to $z_p$ will be off by a factor of
$d_s\left(z_p\right)/d_s\left(z_l\right)$. The surface brightness
dimming correction will be similarly incorrect, with $\mu_p = \mu_l -
10 \log \left[\left(1+z_p\right)/\left(1+z_l\right)\right]$. Finally,
the photoFP fits differ between redshift bins.  A galaxy with an
incorrect photo-z will therefore lie off the photoFP in its assigned
redshift bin by
\begin{equation}\label{eq:dreff_err}
\Delta \log R_e^{\rm{err}} =
\log\left(\frac{d_s\left(z_p\right)}{d_s\left(z_l\right)}
\frac{R_e^p\left(\mu_p,C\right)}{R_e^l\left(\mu_l,C\right)}\right).
\end{equation}
The expressions $R_e^p\left(\mu_p,C\right)$ and
$R_e^l\left(\mu_l,C\right)$ are the radii that would be predicted
by the photoFP for that galaxy's surface brightness and concentration
in the bins corresponding to $z_p$ and $z_l$, respectively.  

The quantity $f_l$ can be estimated by cross-correlating the positions
of sources at $z_p$ with lenses at $z_l$.  We assume that the
positions of galaxies in widely separated redshift bins are
uncorrelated and that any observed excess of sources far behind a lens
is due to scattering from $z_l$.  This means that
\begin{equation}
f_l = \frac{w_{il}(\theta)}{1+w_{il}(\theta)}, 
\end{equation}
where $w_{il}(\theta)$ is the angular cross-correlation between the
positions of sources at $z_i$ and lenses at $z_l$.  A
cross-correlation signal of this form can also be produced by the
boosted number counts of magnified background sources (e.g.,
\citealt{2011MNRAS.411.2113J}) but that effect is too weak to detect
with a lens sample of this size.

The cross-correlations for $0.20 < z_l < 0.25$ with a range of $z_s$
bins are shown in figure \ref{fig:delta_resid}.  We fit an angular
correlation function of the form
\begin{equation}
1 + w_{il}(\theta) = \frac{A_{il}}{\theta^{\,0.8}} + B_{il}
\end{equation} 
where $A_{il}$ and $B_{il}$ are free parameters.  The choice of power
law index is motivated by the angular correlation function
measurements of \citet{2011ApJ...728...46W}, which are in agreement with our observed
$w_{ll}$.  Incorrectly estimating the true mean density of galaxies at
$z_p$ will cause $B_{il}$ to deviate from unity, as is observed.  We
remove the effects of this uncertainty when calculating $f_l$ by
setting $B_{il} = 1$.  Sources with $f_l > 0.20$ (above the black
horizontal line in Figure 2) are excluded from the lensing
measurement, while sources with $f_l < 0.20$ are corrected using
equation \ref{eq:kappa_pzerr}.

In addition to the effects of galaxy clustering on photometric
redshift errors, a mean offset between the true and photometric
redshifts in a $z_p$ bin will cause an incorrect estimation of the
critical density $\Sigma_{\rm{crit}}$ for all of the galaxies in that
bin. This error depends on the distribution of foreground lens
redshifts. Using the method of \citet{2008MNRAS.386..781M}, we
estimate the effect of a mean shift in our photo-z's on the signal of
no more than $10\%$. This uncertainty is small relative to the other
corrections discussed here, so we defer this calculation to Paper II.

\subsection{Sky Proximity Bias Correction}\label{sec:sky_corr}
The SDSS photometric pipeline produces known sky subtraction proximity
effects, where the photometry of objects near bright stars or galaxies
is systematically biased (c.f. \citealt{aihara11}). This may induce a
systematic bias in the estimated radii, surface brightnesses, and
concentrations that contaminates the lensing signal. Sky subtraction
effects cannot distinguish between foreground and background galaxies
(with respect to the bright lens), so this proximity bias can in
principle be estimated from the photoFP residuals for galaxies in the
{\it foreground} of the lenses, which are unaffected by lensing.

\begin{figure}
\includegraphics[width=1.0\linewidth]{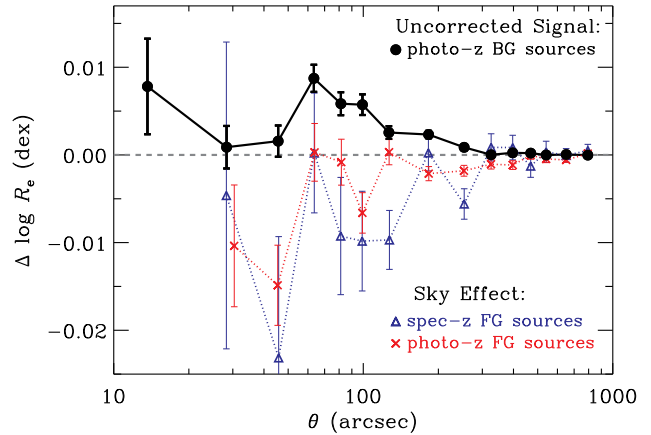}
\caption{The raw magnification signal around the galaxy lenses (filled
  black circles) compared with the sky proximity bias measured from
  foreground sources. The red crosses show the estimated sky
  subtraction effect using sources with photo-z's; the blue triangles
  show the same estimate, but using those foreground galaxies with
  spectroscopic redshifts.  }\label{fig:sky_effects}
\end{figure}

Figure \ref{fig:sky_effects} shows the average deviation from the
photoFP as a function of source-lens angular separation for both
foreground and background sources. The sky proximity bias
systematically induces a reduction in effective radius relative to the
photoFP trend. The lensing signal is thus the difference between the
background and foreground photoFP deviations at each angular
separation. Of note is the fact that our empirical sky correction
extends beyond the size of the SDSS sky subtraction box, which is
$\sim 100''$; this is a result of galaxy clustering. Each of the
bright objects used as a lens will tend to be associated with a galaxy
overdensity on the sky. This excess will also impact the sky
correction, even in neighboring sky subtraction cells, so the angular
scale of the resulting correction will be set by the galaxy
correlation function.

The low redshift of the lens sample and the poor quality of the
photo-z's (which preferentially scatter higher-$z$ sources to lower
$z$) result in a large fraction of source galaxies near lenses with
photometric redshifts $z_p < z_l$ that are actually at $z > z_l$.
This means that a foreground sample of sources with photo-z's will be
contaminated by objects from higher $z$. The cut on $f_L$ described
above removes many such contaminating galaxies, at the cost of
dramatically reducing the signal-to-noise ratio of the sky proximity
bias estimate. This remains the major source of uncertainty in this
measurement.

As a check against this effect, we also show the deviation from the
photoFP trend of those foreground sources with spectroscopic
redshifts. Any large bias to this sky subtraction estimate resulting
from imperfect photo-z's should produce a substantial difference
between the spectroscopic and photometric foreground estimates; this
is not observed.

\subsection{Halo Mass Profile}\label{sec:halo_profile}
\begin{figure*}[t]
\begin{center}
\includegraphics[width=0.7\linewidth]{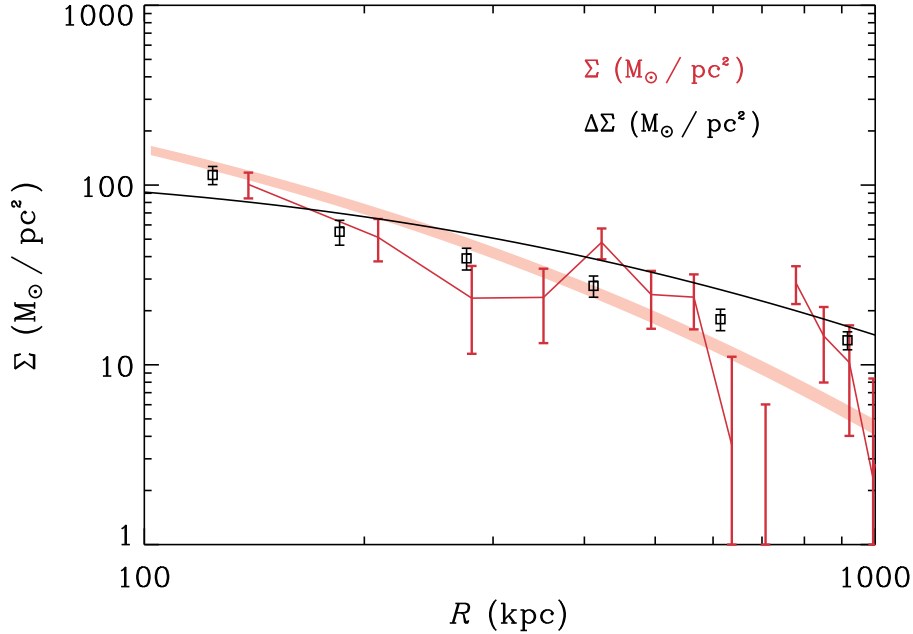}
\caption{Red points: $\Sigma$ from this work.  Black points: $\Delta
  \Sigma$ from M+06 measured using shear.  That measurement used a
  smaller lens sample than we consider here, so we have reduced those
  error bars to allow for a fair comparison with this sample.  The
  solid black line is the best-fit $\Delta \Sigma$ profile from M+06.
  The shaded red region shows the corresponding $\Sigma$ profile (with
  68\% confidence interval) derived from the M+06 data.}
\label{fig:sigma}
\end{center}
\end{figure*}

After controlling for the systematic errors described above, we
calculate the line-of-sight surface matter density $\Sigma$ by
weighting each lens--background source pair by the critical density
for lensing, $\Sigma_{\rm{crit}}\left(z_s,z_l\right)$. We bin this
density by physical separation in the lens plane. Our results are
shown in figure \ref{fig:sigma}, along with existing measurements from
M+06 for a similar lens population.

\section{Discussion: The Way Forward}\label{sec:discussion}
The magnification signal demonstrated above, while many times stronger
than previous magnification measurements, is still somewhat noisier
than the shear signal for a comparable sample. This is because the
convergence dispersion resulting from the measured photoFP width is
35\% (1.8 times larger than the intrinsic shear dispersion of 20\%)
and because we have only used the third of the source sample
consistent with early-type SEDs.

If the fundamental achievable limit for this technique is the
intrinsic scatter in the {\it spectroscopic} fundamental plane, then
the average magnification S/N for an early-type galaxy is the same as
in shear; a comparable photometric Tully-Fisher relation for late-type
galaxies would bring us to the point where magnification and shear
provide comparable information. And any improvement in our
understanding of galaxy evolution and dynamics that further diminishes
the scatter in these scaling relations will boost the magnification
signal beyond that available for shear measurement.

Perhaps just as valuable, magnification by this method is not
sensitive to the same systematic biases that challenge upcoming shear
measurements. For instance, the intrinsic galaxy alignment signal on
large scales should not affect galaxy sizes, concentrations, and mean
surface brightnesses in same the manner in which it affects shapes. We
expect that this technique will also prove useful in extracting and
removing instrumental sytematics, such as those arising from
variations in the telescope point-spread function, and will
investigate this prospect in a subsequent paper.

\acknowledgements

The authors are deeply grateful to Rachel Mandelbaum and Reiko
Nakajima for their help in understanding several of the systematic
errors, and to them and to Chris Hirata for many useful and productive
discussions related to this work. The authors are also grateful to the
PRIMUS team for allowing the use of their spectroscopic catalog in
estimating photometric redshift calibration errors.

E.~M.~H. is supported by Award \#DE-AC02-05CH11231, funded by the
Department of Energy's Office of High Energy Physics. G.~G. is
supported by a fellowship from the Miller Institute for Basic Research
in Science.

Funding for SDSS-III has been provided by the Alfred P. Sloan
Foundation, the Participating Institutions, the National Science
Foundation, and the U.S. Department of Energy. The SDSS-III web site
is http://www.sdss3.org/.

SDSS-III is managed by the Astrophysical Research Consortium. The
Participating Institutions are the American Museum of Natural History,
Astrophysical Institute Potsdam, University of Basel, University of
Cambridge, Case Western Reserve University, University of Chicago,
Drexel University, Fermilab, the Institute for Advanced Study, the
Japan Participation Group, Johns Hopkins University, the Joint
Institute for Nuclear Astrophysics, the Kavli Institute for Particle
Astrophysics and Cosmology, the Korean Scientist Group, the Chinese
Academy of Sciences (LAMOST), Los Alamos National Laboratory, the
Max-Planck-Institute for Astronomy (MPIA), the Max-Planck-Institute
for Astrophysics (MPA), New Mexico State University, Ohio State
University, University of Pittsburgh, University of Portsmouth,
Princeton University, the United States Naval Observatory, and the
University of Washington.


\begin{thebibliography}{50}
\expandafter\ifx\csname natexlab\endcsname\relax\def\natexlab#1{#1}\fi

\bibitem[{{Abazajian} {et~al.}(2009){Abazajian}, {Adelman-McCarthy},
  {Ag{\"u}eros}, {Allam}, {Allende Prieto}, {An}, {Anderson}, {Anderson},
  {Annis}, {Bahcall}, \& et~al.}]{abazajian09_dr7}
{Abazajian}, K.~N., {et~al.} 2009, \apjs, 182, 543

\bibitem[{{Aihara} {et~al.}(2011){Aihara}, {Allende Prieto}, {An}, {Anderson},
  {Aubourg}, {Balbinot}, {Beers}, {Berlind}, {Bickerton}, {Bizyaev}, {Blanton},
  {Bochanski}, {Bolton}, {Bovy}, {Brandt}, {Brinkmann}, {Brown}, {Brownstein},
  {Busca}, {Campbell}, {Carr}, {Chen}, {Chiappini}, {Comparat}, {Connolly},
  {Cortes}, {Croft}, {Cuesta}, {da Costa}, {Davenport}, {Dawson}, {Dhital},
  {Ealet}, {Ebelke}, {Edmondson}, {Eisenstein}, {Escoffier}, {Esposito},
  {Evans}, {Fan}, {Femen{\'{\i}}a Castell{\'a}}, {Font-Ribera}, {Frinchaboy},
  {Ge}, {Gillespie}, {Gilmore}, {Gonz{\'a}lez Hern{\'a}ndez}, {Gott}, {Gould},
  {Grebel}, {Gunn}, {Hamilton}, {Harding}, {Harris}, {Hawley}, {Hearty}, {Ho},
  {Hogg}, {Holtzman}, {Honscheid}, {Inada}, {Ivans}, {Jiang}, {Johnson},
  {Jordan}, {Jordan}, {Kazin}, {Kirkby}, {Klaene}, {Knapp}, {Kneib},
  {Kochanek}, {Koesterke}, {Kollmeier}, {Kron}, {Lampeitl}, {Lang}, {Le Goff},
  {Lee}, {Lin}, {Long}, {Loomis}, {Lucatello}, {Lundgren}, {Lupton}, {Ma},
  {MacDonald}, {Mahadevan}, {Maia}, {Makler}, {Malanushenko}, {Malanushenko},
  {Mandelbaum}, {Maraston}, {Margala}, {Masters}, {McBride}, {McGehee},
  {McGreer}, {M{\'e}nard}, {Miralda-Escud{\'e}}, {Morrison}, {Mullally},
  {Muna}, {Munn}, {Murayama}, {Myers}, {Naugle}, {Fausti Neto}, {Cuong Nguyen},
  {Nichol}, {O'Connell}, {Ogando}, {Olmstead}, {Oravetz}, {Padmanabhan},
  {Palanque-Delabrouille}, {Pan}, {Pandey}, {P{\^a}ris}, {Percival},
  {Petitjean}, {Pfaffenberger}, {Pforr}, {Phleps}, {Pichon}, {Pieri}, {Prada},
  {Price-Whelan}, {Raddick}, {Ramos}, {Reyl{\'e}}, {Rich}, {Richards}, {Rix},
  {Robin}, {Rocha-Pinto}, {Rockosi}, {Roe}, {Rollinde}, {Ross}, {Ross},
  {Rossetto}, {S{\'a}nchez}, {Sayres}, {Schlegel}, {Schlesinger}, {Schmidt},
  {Schneider}, {Sheldon}, {Shu}, {Simmerer}, {Simmons}, {Sivarani}, {Snedden},
s  {Sobeck}, {Steinmetz}, {Strauss}, {Szalay}, {Tanaka}, {Thakar}, {Thomas},
  {Tinker}, {Tofflemire}, {Tojeiro}, {Tremonti}, {Vandenberg}, {Vargas
  Maga{\~n}a}, {Verde}, {Vogt}, {Wake}, {Wang}, {Weaver}, {Weinberg}, {White},
  {White}, {Yanny}, {Yasuda}, {Yeche}, \& {Zehavi}}]{aihara11}
{Aihara}, H., {et~al.} 2011, \apjs, 193, 29

\bibitem[{{Bauer} {et~al.}(2011){Bauer}, {Seitz}, {Jerke}, {Scalzo},
  {Rabinowitz}, {Ellman}, \& {Baltay}}]{bauer11}
{Bauer}, A.~H., {Seitz}, S., {Jerke}, J., {Scalzo}, R., {Rabinowitz}, D.,
  {Ellman}, N., \& {Baltay}, C. 2011, \apj, 732, 64

\bibitem[{{Bell} {et~al.}(2003){Bell}, {McIntosh}, {Katz}, \&
  {Weinberg}}]{bell03}
{Bell}, E.~F., {McIntosh}, D.~H., {Katz}, N., \& {Weinberg}, M.~D. 2003, \apjs,
  149, 289

\bibitem[{{Berg{\'e}} {et~al.}(2008){Berg{\'e}}, {Pacaud}, {R{\'e}fr{\'e}gier},
  {Massey}, {Pierre}, {Amara}, {Birkinshaw}, {Paulin-Henriksson}, {Smith}, \&
  {Willis}}]{berge08}
{Berg{\'e}}, J., {et~al.} 2008, \mnras, 385, 695

\bibitem[{{Bernardi} {et~al.}(2003){Bernardi}, {Sheth}, {Annis}, {Burles},
  {Eisenstein}, {Finkbeiner}, {Hogg}, {Lupton}, {Schlegel}, {SubbaRao},
  {Bahcall}, {Blakeslee}, {Brinkmann}, {Castander}, {Connolly}, {Csabai},
  {Doi}, {Fukugita}, {Frieman}, {Heckman}, {Hennessy}, {Ivezi{\'c}}, {Knapp},
  {Lamb}, {McKay}, {Munn}, {Nichol}, {Okamura}, {Schneider}, {Thakar}, \&
  {York}}]{bernardi03c}
{Bernardi}, M., {et~al.} 2003, \aj, 125, 1866

\bibitem[{{Bertin} \& {Lombardi}(2006)}]{bertin06}
{Bertin}, G., \& {Lombardi}, M. 2006, \apjl, 648, L17

\bibitem[{{Blanton} {et~al.}(2005){Blanton}, {Schlegel}, {Strauss},
  {Brinkmann}, {Finkbeiner}, {Fukugita}, {Gunn}, {Hogg}, {Ivezi{\'c}}, {Knapp},
  {Lupton}, {Munn}, {Schneider}, {Tegmark}, \& {Zehavi}}]{blanton05_vagc}
{Blanton}, M.~R., {et~al.} 2005, \aj, 129, 2562

\bibitem[{{Brada{\v c}} {et~al.}(2008){Brada{\v c}}, {Schrabback}, {Erben},
  {McCourt}, {Million}, {Mantz}, {Allen}, {Blandford}, {Halkola},
  {Hildebrandt}, {Lombardi}, {Marshall}, {Schneider}, {Treu}, \&
  {Kneib}}]{bradac08}
{Brada{\v c}}, M., {et~al.} 2008, \apj, 681, 187

\bibitem[{{Broadhurst} {et~al.}(2005){Broadhurst}, {Takada}, {Umetsu}, {Kong},
  {Arimoto}, {Chiba}, \& {Futamase}}]{broadhurst05}
{Broadhurst}, T., {Takada}, M., {Umetsu}, K., {Kong}, X., {Arimoto}, N.,
  {Chiba}, M., \& {Futamase}, T. 2005, \apjl, 619, L143

\bibitem[{{Clowe} {et~al.}(2006){Clowe}, {Brada{\v c}}, {Gonzalez},
  {Markevitch}, {Randall}, {Jones}, \& {Zaritsky}}]{clowe06_bullet}
{Clowe}, D., {Brada{\v c}}, M., {Gonzalez}, A.~H., {Markevitch}, M., {Randall},
  S.~W., {Jones}, C., \& {Zaritsky}, D. 2006, \apjl, 648, L109

\bibitem[{{Davis} {et~al.}(2003){Davis}, {Faber}, {Newman}, {Phillips},
  {Ellis}, {Steidel}, {Conselice}, {Coil}, {Finkbeiner}, {Koo}, {Guhathakurta},
  {Weiner}, {Schiavon}, {Willmer}, {Kaiser}, {Luppino}, {Wirth}, {Connolly},
  {Eisenhardt}, {Cooper}, \& {Gerke}}]{2003SPIE.4834..161D}
{Davis}, M., {et~al.} 2003, in Presented at the Society of Photo-Optical
  Instrumentation Engineers (SPIE) Conference, Vol. 4834, Society of
  Photo-Optical Instrumentation Engineers (SPIE) Conference Series, ed.
  {P.~Guhathakurta}, 161--172

\bibitem[{{Djorgovski} \& {Davis}(1987)}]{djorgovski87}
{Djorgovski}, S., \& {Davis}, M. 1987, \apj, 313, 59

\bibitem[{{Dressler} {et~al.}(1987){Dressler}, {Lynden-Bell}, {Burstein},
  {Davies}, {Faber}, {Terlevich}, \& {Wegner}}]{dressler87}
{Dressler}, A., {Lynden-Bell}, D., {Burstein}, D., {Davies}, R.~L., {Faber},
  S.~M., {Terlevich}, R., \& {Wegner}, G. 1987, \apj, 313, 42

\bibitem[{{Eisenstein} {et~al.}(2003){Eisenstein}, {Hogg}, {Fukugita},
  {Nakamura}, {Bernardi}, {Finkbeiner}, {Schlegel}, {Brinkmann}, {Connolly},
  {Csabai}, {Gunn}, {Ivezi{\'c}}, {Lamb}, {Loveday}, {Munn}, {Nichol},
  {Schneider}, {Strauss}, {Szalay}, \& {York}}]{eisenstein03}
{Eisenstein}, D.~J., {et~al.} 2003, \apj, 585, 694

\bibitem[{{Feldmann} {et~al.}(2006){Feldmann}, {Carollo}, {Porciani}, {Lilly},
  {Capak}, {Taniguchi}, {Le F{\`e}vre}, {Renzini}, {Scoville}, {Ajiki},
  {Aussel}, {Contini}, {McCracken}, {Mobasher}, {Murayama}, {Sanders},
  {Sasaki}, {Scarlata}, {Scodeggio}, {Shioya}, {Silverman}, {Takahashi},
  {Thompson}, \& {Zamorani}}]{feldmann06}
{Feldmann}, R., {et~al.} 2006, \mnras, 372, 565

\bibitem[{{Fischer} {et~al.}(2000){Fischer}, {McKay}, {Sheldon}, {Connolly},
  {Stebbins}, {Frieman}, {Jain}, {Joffre}, {Johnston}, {Bernstein}, {Annis},
  {Bahcall}, {Brinkmann}, {Carr}, {Csabai}, {Gunn}, {Hennessy}, {Hindsley},
  {Hull}, {Ivezi{\'c}}, {Knapp}, {Limmongkol}, {Lupton}, {Munn}, {Nash},
  {Newberg}, {Owen}, {Pier}, {Rockosi}, {Schneider}, {Smith}, {Stoughton},
  {Szalay}, {Szokoly}, {Thakar}, {Vogeley}, {Waddell}, {Weinberg}, {York}, \&
  {The SDSS Collaboration}}]{fischer00}
{Fischer}, P., {et~al.} 2000, \aj, 120, 1198

\bibitem[{{Fu} {et~al.}(2008){Fu}, {Semboloni}, {Hoekstra}, {Kilbinger}, {van
  Waerbeke}, {Tereno}, {Mellier}, {Heymans}, {Coupon}, {Benabed}, {Benjamin},
  {Bertin}, {Dor{\'e}}, {Hudson}, {Ilbert}, {Maoli}, {Marmo}, {McCracken}, \&
  {M{\'e}nard}}]{fu08}
{Fu}, L., {et~al.} 2008, \aap, 479, 9

\bibitem[{{Graham}(2002)}]{graham02}
{Graham}, A.~W. 2002, \mnras, 334, 859

\bibitem[{{Graves} \& {Faber}(2010)}]{graves10_paperIII}
{Graves}, G.~J., \& {Faber}, S.~M. 2010, \apj, 717, 803

\bibitem[{{Hildebrandt} {et~al.}(2011){Hildebrandt}, {Muzzin}, {Erben},
  {Hoekstra}, {Kuijken}, {Surace}, {van Waerbeke}, {Wilson}, \&
  {Yee}}]{hildebrandt11}
{Hildebrandt}, H., {et~al.} 2011, \apj, 733, 30

\bibitem[{{Hoekstra}(2007)}]{hoekstra07}
{Hoekstra}, H. 2007, \mnras, 379, 317

\bibitem[{{Hoekstra} {et~al.}(2004){Hoekstra}, {Yee}, \&
  {Gladders}}]{hoekstra04}
{Hoekstra}, H., {Yee}, H.~K.~C., \& {Gladders}, M.~D. 2004, \apj, 606, 67

\bibitem[{{Jain} \& {Lima}(2011)}]{2011MNRAS.411.2113J}
{Jain}, B., \& {Lima}, M. 2011, \mnras, 411, 2113

\bibitem[{{Jee} {et~al.}(2007){Jee}, {Ford}, {Illingworth}, {White},
  {Broadhurst}, {Coe}, {Meurer}, {van der Wel}, {Ben{\'{\i}}tez}, {Blakeslee},
  {Bouwens}, {Bradley}, {Demarco}, {Homeier}, {Martel}, \& {Mei}}]{jee07}
{Jee}, M.~J., {et~al.} 2007, \apj, 661, 728

\bibitem[{{J{\o}rgensen} {et~al.}(1996){J{\o}rgensen}, {Franx}, \&
  {Kjaergaard}}]{jorgensen96}
{J{\o}rgensen}, I., {Franx}, M., \& {Kjaergaard}, P. 1996, \mnras, 280, 167

\bibitem[{{Kneib} {et~al.}(2003){Kneib}, {Hudelot}, {Ellis}, {Treu}, {Smith},
  {Marshall}, {Czoske}, {Smail}, \& {Natarajan}}]{kneib03}
{Kneib}, J., {et~al.} 2003, \apj, 598, 804

\bibitem[{{Kubo} {et~al.}(2009){Kubo}, {Annis}, {Hardin}, {Kubik}, {Lawhorn},
  {Lin}, {Nicklaus}, {Nelson}, {Reis}, {Seo}, {Soares-Santos}, {Stebbins}, \&
  {Yunker}}]{kubo09}
{Kubo}, J.~M., {et~al.} 2009, \apjl, 702, L110

\bibitem[{{Le F{\`e}vre} {et~al.}(2005){Le F{\`e}vre}, {Vettolani}, {Garilli},
  {Tresse}, {Bottini}, {Le Brun}, {Maccagni}, {Picat}, {Scaramella},
  {Scodeggio}, {Zanichelli}, {Adami}, {Arnaboldi}, {Arnouts}, {Bardelli},
  {Bolzonella}, {Cappi}, {Charlot}, {Ciliegi}, {Contini}, {Foucaud},
  {Franzetti}, {Gavignaud}, {Guzzo}, {Ilbert}, {Iovino}, {McCracken}, {Marano},
  {Marinoni}, {Mathez}, {Mazure}, {Meneux}, {Merighi}, {Paltani}, {Pell{\`o}},
  {Pollo}, {Pozzetti}, {Radovich}, {Zamorani}, {Zucca}, {Bondi}, {Bongiorno},
  {Busarello}, {Lamareille}, {Mellier}, {Merluzzi}, {Ripepi}, \&
  {Rizzo}}]{2005A&A...439..845L}
{Le F{\`e}vre}, O., {et~al.} 2005, \aap, 439, 845

\bibitem[{{Leauthaud} {et~al.}(2010){Leauthaud}, {Finoguenov}, {Kneib},
  {Taylor}, {Massey}, {Rhodes}, {Ilbert}, {Bundy}, {Tinker}, {George}, {Capak},
  {Koekemoer}, {Johnston}, {Zhang}, {Cappelluti}, {Ellis}, {Elvis}, {Giodini},
  {Heymans}, {Le F{\`e}vre}, {Lilly}, {McCracken}, {Mellier},
  {R{\'e}fr{\'e}gier}, {Salvato}, {Scoville}, {Smoot}, {Tanaka}, {Van
  Waerbeke}, \& {Wolk}}]{leauthaud10}
{Leauthaud}, A., {et~al.} 2010, \apj, 709, 97

\bibitem[{{Lima} {et~al.}(2008){Lima}, {Cunha}, {Oyaizu}, {Frieman}, {Lin}, \&
  {Sheldon}}]{2008MNRAS.390..118L}
{Lima}, M., {Cunha}, C.~E., {Oyaizu}, H., {Frieman}, J., {Lin}, H., \&
  {Sheldon}, E.~S. 2008, \mnras, 390, 118

\bibitem[{{Mahdavi} {et~al.}(2007{\natexlab{a}}){Mahdavi}, {Hoekstra}, {Babul},
  {Balam}, \& {Capak}}]{mahdavi07_a520}
{Mahdavi}, A., {Hoekstra}, H., {Babul}, A., {Balam}, D.~D., \& {Capak}, P.~L.
  2007{\natexlab{a}}, \apj, 668, 806

\bibitem[{{Mahdavi} {et~al.}(2007{\natexlab{b}}){Mahdavi}, {Hoekstra}, {Babul},
  {Sievers}, {Myers}, \& {Henry}}]{mahdavi07_a478}
{Mahdavi}, A., {Hoekstra}, H., {Babul}, A., {Sievers}, J., {Myers}, S.~T., \&
  {Henry}, J.~P. 2007{\natexlab{b}}, \apj, 664, 162

\bibitem[{{Mandelbaum} {et~al.}(2006{\natexlab{a}}){Mandelbaum}, {Seljak},
  {Cool}, {Blanton}, {Hirata}, \& {Brinkmann}}]{mandelbaum06_massprofiles}
{Mandelbaum}, R., {Seljak}, U., {Cool}, R.~J., {Blanton}, M., {Hirata}, C.~M.,
  \& {Brinkmann}, J. 2006{\natexlab{a}}, \mnras, 372, 758

\bibitem[{{Mandelbaum} {et~al.}(2008{\natexlab{a}}){Mandelbaum}, {Seljak}, \&
  {Hirata}}]{mandelbaum08}
{Mandelbaum}, R., {Seljak}, U., \& {Hirata}, C.~M. 2008{\natexlab{a}}, JCAP, 8,
  6

\bibitem[{{Mandelbaum} {et~al.}(2006{\natexlab{b}}){Mandelbaum}, {Seljak},
  {Kauffmann}, {Hirata}, \& {Brinkmann}}]{mandelbaum06_halomass}
{Mandelbaum}, R., {Seljak}, U., {Kauffmann}, G., {Hirata}, C.~M., \&
  {Brinkmann}, J. 2006{\natexlab{b}}, \mnras, 368, 715

\bibitem[{{Mandelbaum} {et~al.}(2005){Mandelbaum}, {Hirata}, {Seljak}, {Guzik},
  {Padmanabhan}, {Blake}, {Blanton}, {Lupton}, \&
  {Brinkmann}}]{2005MNRAS.361.1287M}
{Mandelbaum}, R., {et~al.} 2005, \mnras, 361, 1287

\bibitem[{{Mandelbaum} {et~al.}(2008{\natexlab{b}}){Mandelbaum}, {Seljak},
  {Hirata}, {Bardelli}, {Bolzonella}, {Bongiorno}, {Carollo}, {Contini},
  {Cunha}, {Garilli}, {Iovino}, {Kampczyk}, {Kneib}, {Knobel}, {Koo},
  {Lamareille}, {Le F{\`e}vre}, {Le Borgne}, {Lilly}, {Maier}, {Mainieri},
  {Mignoli}, {Newman}, {Oesch}, {Perez-Montero}, {Ricciardelli}, {Scodeggio},
  {Silverman}, \& {Tasca}}]{2008MNRAS.386..781M}
---. 2008{\natexlab{b}}, \mnras, 386, 781

\bibitem[{{Massey} {et~al.}(2007){Massey}, {Rhodes}, {Leauthaud}, {Capak},
  {Ellis}, {Koekemoer}, {R{\'e}fr{\'e}gier}, {Scoville}, {Taylor}, {Albert},
  {Berg{\'e}}, {Heymans}, {Johnston}, {Kneib}, {Mellier}, {Mobasher},
  {Semboloni}, {Shopbell}, {Tasca}, \& {Van Waerbeke}}]{2007ApJS..172..239M}
{Massey}, R., {et~al.} 2007, \apjs, 172, 239

\bibitem[{{M{\'e}nard} {et~al.}(2010){M{\'e}nard}, {Scranton}, {Fukugita}, \&
  {Richards}}]{2010MNRAS.405.1025M}
{M{\'e}nard}, B., {Scranton}, R., {Fukugita}, M., \& {Richards}, G. 2010,
  \mnras, 405, 1025

\bibitem[{{Okabe} \& {Umetsu}(2008)}]{okabe08}
{Okabe}, N., \& {Umetsu}, K. 2008, \pasj, 60, 345

\bibitem[{{Parker} {et~al.}(2007){Parker}, {Hoekstra}, {Hudson}, {van
  Waerbeke}, \& {Mellier}}]{parker07}
{Parker}, L.~C., {Hoekstra}, H., {Hudson}, M.~J., {van Waerbeke}, L., \&
  {Mellier}, Y. 2007, \apj, 669, 21

\bibitem[{{Rozo} \& {Schmidt}(2010)}]{2010arXiv1009.5735R}
{Rozo}, E., \& {Schmidt}, F. 2010, ArXiv e-prints

\bibitem[{{Scranton} {et~al.}(2005){Scranton}, {M{\'e}nard}, {Richards},
  {Nichol}, {Myers}, {Jain}, {Gray}, {Bartelmann}, {Brunner}, {Connolly},
  {Gunn}, {Sheth}, {Bahcall}, {Brinkman}, {Loveday}, {Schneider}, {Thakar}, \&
  {York}}]{2005ApJ...633..589S}
{Scranton}, R., {et~al.} 2005, \apj, 633, 589

\bibitem[{{Seljak} {et~al.}(2005){Seljak}, {Makarov}, {Mandelbaum}, {Hirata},
  {Padmanabhan}, {McDonald}, {Blanton}, {Tegmark}, {Bahcall}, \&
  {Brinkmann}}]{seljak05}
{Seljak}, U., {et~al.} 2005, \prd, 71, 043511

\bibitem[{{Sheldon} {et~al.}(2004){Sheldon}, {Johnston}, {Frieman}, {Scranton},
  {McKay}, {Connolly}, {Budav{\'a}ri}, {Zehavi}, {Bahcall}, {Brinkmann}, \&
  {Fukugita}}]{sheldon04}
{Sheldon}, E.~S., {et~al.} 2004, \aj, 127, 2544

\bibitem[{{Sheldon} {et~al.}(2009){Sheldon}, {Johnston}, {Scranton}, {Koester},
  {McKay}, {Oyaizu}, {Cunha}, {Lima}, {Lin}, {Frieman}, {Wechsler}, {Annis},
  {Mandelbaum}, {Bahcall}, \& {Fukugita}}]{sheldon09}
---. 2009, \apj, 703, 2217

\bibitem[{{Shimasaku} {et~al.}(2001){Shimasaku}, {Fukugita}, {Doi}, {Hamabe},
  {Ichikawa}, {Okamura}, {Sekiguchi}, {Yasuda}, {Brinkmann}, {Csabai},
  {Ichikawa}, {Ivezi{\'c}}, {Kunszt}, {Schneider}, {Szokoly}, {Watanabe}, \&
  {York}}]{shimasaku01}
{Shimasaku}, K., {et~al.} 2001, \aj, 122, 1238

\bibitem[Wake et al.(2011)]{2011ApJ...728...46W} Wake, D.~A., et al.\ 2011, 
\apj, 728, 46 




\end{thebibliography}

\end{document}